\documentclass[10pt]{article}
\usepackage{epsfig}
\renewcommand{\author}[2]{\begin{center}
                           {\sc #1}\\
                           {#2}
                          \end{center}}
\renewcommand{\title}[1]{\begin{center}
                      {\Large {\bf #1}}
                      \end{center}}

 \def\ie{{\sl i.e.,}\,}
 \def\eg{{\sl e.g.,}\,}

 \def\la{\hbox{\raise.5ex\hbox{$<$}
     \kern-1.1em\lower.5ex\hbox{$\sim$}}}
 \def\ga{\hbox{\raise.5ex\hbox{$>$}
     \kern-1.1em\lower.5ex\hbox{$\sim$}}}

\def\apj{{\it ApJ}}			
\def\apjl{{\it ApJ}}		
\def\apjs{{\it ApJS}}

\def\mnras{{\it MNRAS}}

\begin{document}

\title{Simulations of Relativistic Jets with GENESIS
\vspace*{5mm}}
\author{M.A. Aloy$^1$, 
        J.M$^{\underline{\mbox{a}}}$ Ib\'a\~nez$^1$,
        J.M$^{\underline{\mbox{a}}}$ Mart\'{\i}$^1$ 
        J.L. G\'omez$^2$  and
        E. M\"uller$^3$
\vspace{5mm}}

\begin{center}
\small{$^1$Departamento de Astronom\'{\i}a y Astrof\'{\i}sica,
                  UVEG, 46100 Burjassot,
                  Spain. \\ e-mails:~Miguel.A.Aloy@uv.es,
                  Jose.M.Ibanez@uv.es, Jose.M.Marti@uv.es}
\vspace{4mm}

\noindent
\small{$^3$Instituto de Astrof\'{\i}sica de Andaluc\'{\i}a (CSIC)
                  Granada, Spain. \\ e-mail:~jlgomez@iaa.es}
\vspace{4mm}

\noindent
\small{$^3$Max-Planck-Institut f\"ur Astrophysik,
           85748 Garching, Germany. \\ e-mail:~ewald@mpa-garching.mpg.de}

\end{center}

\vspace{5mm} 

\begin{abstract}
 The multidimensional relativistic hydrodynamical code GENESIS has been 
used to obtain first results of {\it 3D} simulations of relativistic jets.
We have studied the influence of a slight perturbation of the injection 
velocity field on the morphodynamics of otherwise axisymmetric 
relativistic jets.
\end{abstract}

\section{Introduction } \label{s:intro}

  Astrophysical jets are continuous channels of plasma produced by
some active galactic nuclei that are currently observed in radio
frequences. The relativistic nature of the plasma has been inferred
from (esentially) two observational evidences: (i) the existence of
superluminal motions of some radio components and, (ii) the high flux
variability (even smaller than one day for some sources). Since
seve\-ral years the dynamical and morphological properties of
axisymmetric relativistic jets are investigated by means of
relativistic hydrodynamic simula\-tions (\eg \cite{Putten93, DH94,
Metal97, KF98}).  In addition, relativistic MHD simulations have been
performed in 2D (\cite{KNM96, Koide97}) and 3D (\cite{Netal97,
Netal98}). In their 3D simulations, \cite{Netal97} and \cite{Netal98},
have studied mildly relativistic jets (Lorentz factor, $W = 4.56$)
propagating both along and obliquely to an ambient magnetic field.

  In this work we report on high-resolution 3D simulations of
relativistic jets with the largest beam flow Lorentz factor performed
up to now (7.09), the largest resolution (8 cells per beam radius),
and covering the longest time evolution (75 normalized time units; a
normalized time unit is defined as the time needed for the jet to
cross a unit length). These facts together with the high performance 
of our hydrodynamic code allowed us to study the morphology and dynamics 
of 3D relativistic jets for the first time.

  The calculations have been performed with the high--resolution 3D
relativistic hydrodynamics code GENESIS \cite{AIMM99a}, which is
an upgraded version of the code developed by \cite{Metal97}.
GENESIS integrates the 3D relativistic hydrodynamic equations in conservation 
form in Cartesian coordinates including an additional conservation equation
for the beam-to-external density fraction to distinguish between beam
and external medium fluids.  
The code is based on a {\it method of lines} which first discretizes the 
spatial part of the relativistic Euler equations and solves the fluxes using 
the Marquina's flux formula \cite{Detal98}. Then the semidiscrete system of 
ordinary differential equations is solved using a third order Runge-Kutta 
algorithm \cite{SO89}. High spatial accuracy is achieved by means of a PPM 
third order interpolation \cite{CW84}. The computations were performed on a
Cartesian domain (X,Y,Z) of size $15R_b \times 15R_b \times 75R_b$ 
($120 \times 120 \times 600$ computational cells), where $R_b$ is the 
beam radius. The jet is injected at $z=0$ in the direction of the 
positive $z$-axis through a circular nozzle defined by $x^2+y^2 \le R_b^2$. 
Beam material is injected with a beam mass fraction $f=1$, and the 
computational domain is initially filled with an external medium ($f=0$). 

  We have considered a 3D model corresponding to model C2 of
\cite{Metal97}, which is characterized by a beam-to-external proper
rest-mass density ratio $\eta=0.01$, a beam Mach number $M_b=6.0$, and
a beam flow speed $v_b = 0.99c$ ($c$ is the speed of light) or a beam
Lorentz factor $W_b \approx 7.09$. An ideal gas equation of state with
an adiabatic exponent $\gamma =5/3$ describes both the jet matter and
the ambient gas. The beam is assumed to be in pressure equilibrium
with the ambient medium.

  The evolution of the jet was simulated up to $T \approx 150 R_b/c$,
when the head of the jet is about to leave the grid. The scaled final time $T \approx 4.6\,10^4 \left(R_b/100\, {\rm pc}\right)\, {\rm yr}$ is about two 
orders of magnitude smaller than the estimated ages of powerful jets.  
Hence, our simulations cannot describe the long term evolution of these 
sources.
  
  Non--axisymmetry was imposed by means of a helical velocity
perturbation at the nozzle given by
\begin{equation}
  v^x_b = \zeta v_b \cos \left(\frac{2 \pi t}{\tau}\right), \;
  v^y_b = \zeta v_b \sin \left(\frac{2 \pi t}{\tau}\right), \;
  v^z_b = v_b \sqrt{1-\zeta^2},
\end{equation}
where $\zeta$ is the ratio of the toroidal to total velocity and
$\tau$ the perturbation period (\ie $\tau = T/n$, $n$ being the number
of cycles completed during the whole simulation).  This velocity field
causes a differential rotation of the beam. The perturbation is chosen
such that it does not change the velocity modulus, (\ie mass, momentum
and energy fluxes of the beam are preserved).

\section{Morphodynamics of 3D relativistic jets}
\label{s:morpho}

\begin{figure}
\epsfig{file=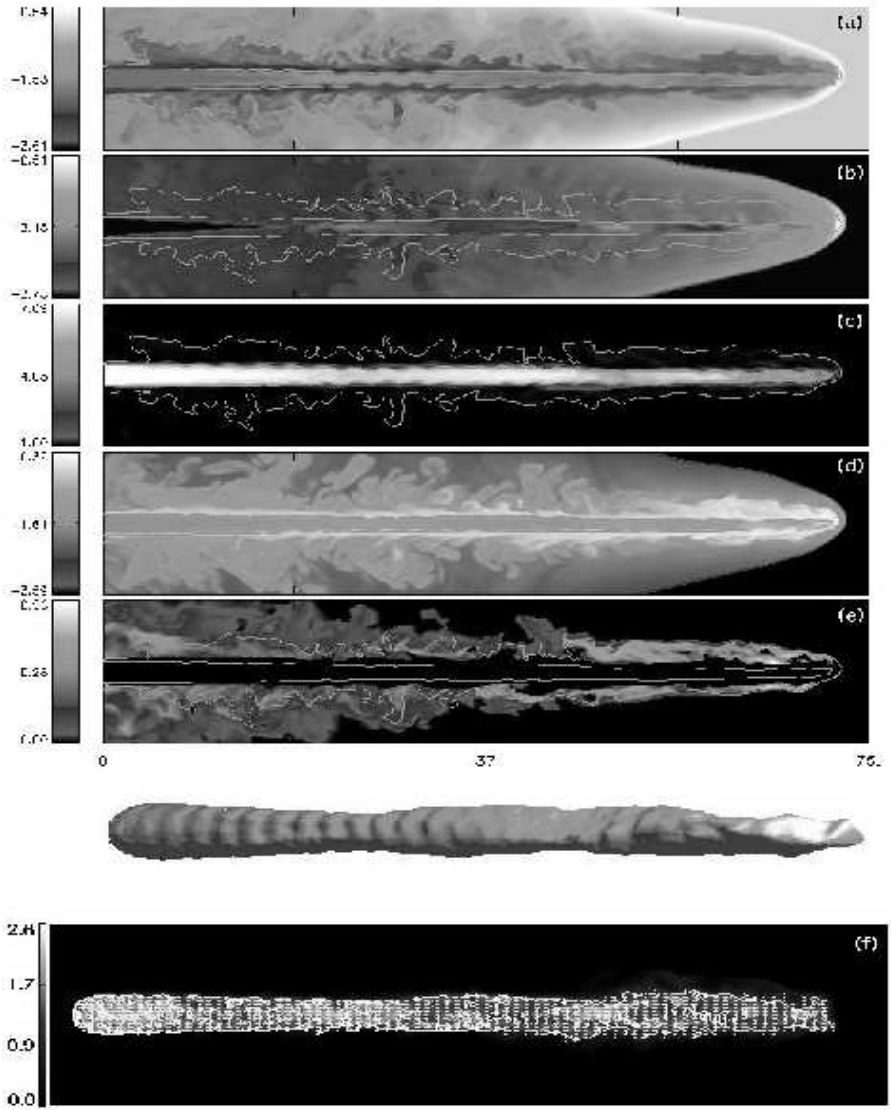,width=12.2cm,height=15.2cm}
\caption{\small Rest-mass density, pressure, flow Lorentz factor, specific
internal energy and backflow velocity distributions (from top to
bottom) of the model discussed in the text in the plane $y=0$ at the
end of the simulation.  White contour levels appearing in each frame
correspond to values of $f$ equal to 0.95 (inner contour; representative of 
the beam) and 0.05 (representative of the cocoon/shocked external medium
interface). The panel under (e) displays the isosurface of $f=0.95$. Panel (f):
Mean local apparent speed observed at an angle of 40 degrees. Arrows show the 
projected direction and magnitude of the apparent motion the contours 
corresponding to values of 1.0\,$c$, 1.6\,$c$, and 2.2\,$c$, respectively. 
Averages have been computed along each line of sight using the emission 
coefficient as a weight.
\label{f:Aa}}
\end{figure}

  Here we consider two models: A, which has a $1\%$ perturbation in 
helical velocity ($\zeta=0.01$) and $n=50$ and B, with $\zeta=0.05$ and
$n=15$. Figure\,\ref{f:Aa} shows various quantities of the model A
in the plane $y=0$ at the end of the simulation. Two values of 
the beam mass fraction are marked by white contour levels. The beam structure
is dominated by the imposed helical pattern with amplitudes of 
$\approx 0.2 R_b$ and $\approx 1.2 R_b$ for A and B, respectively. 

  The overall jet's morphology is characterized by the presence of a highly
turbulent, subsonic cocoon. The pressure distribution outside the beam is 
nearly homogeneous giving rise to a symmetric bow shock 
(Fig.\,\ref{f:Aa}b) in model A. Model B shows a very 
inhomogeneous pressure distribution in the cocoon. As in the classical case 
\cite{Norman96}, the relativistic 3D simulation shows less ordered 
structures in the cocoon than the axisymmetric models. As seen from the 
beam mass fraction levels, the cocoon remains quite thin 
($\sim 2 R_b$) in A and widens ($\sim 4 R_b$) in B.

  The flow field outside the beam shows that the high velocity backflow is
restric\-ted to a small region in the vicinity of the hot spot
(Fig.\,\ref{f:Aa}e), the largest backflow velocities ($\sim
0.5c$) being significantly smaller than in 2D models. The flow annulus
with high Lorentz factor found in axisymmetric simulations is also present, 
but it is reduced to a thin layer around the beam and possesses
sub-relativistic speeds ($\sim 0.25c$) in model A and mildly relativistic
($\sim 0.7$) in B. The size of the backflow 
velocities in the cocoon do not support relativistic beaming in case of 
small perturbations but such possibility is open in larger ones.

  Within the beam the perturbation pattern is superimposed to the
conical shocks at about 26 and 50\,$R_b$. The beam of A does not exhibit
the strong perturbations (deflection, twisting, flattening or even
filamentation) found by other authors \cite{Norman96} for 3D classical
hydrodynamic jets; \cite{Hardee96} for 3D classical MHD jets). This can
be taken as a sign of stability, although it can be argued that our
simulation is not evolved far enough.  For n15p01, the beam is about to
be disrupted at the end of our simulation. Obviously, the beam cross
section and the internal conical shock structure are correlated
(Figure\,\ref{f:Aa}). 

  The helical pattern propagates along the jet at nearly the beam
speed which could yield to superluminal components when viewed at
appropriate angles. Besides this superluminal pattern, the presence of
emitting fluid elements moving at different velocities and
orientations could lead to local variations of the apparent
superluminal motion within the jet.  This is shown in
Fig.\,\ref{f:Aa}f, where we have computed the mean (along each
line of sight, and for a viewing angle of 40 degrees) local apparent
speed. The distribution of apparent motions is inhomogeneous and
resembles that of the observed individual features within knots in M87
\cite{BZO95}.

  The jet can be traced continuously up to the hot spot which
propagates as a strong shock through the ambient medium. Beam material
impinges on the hot spot at high Lorentz factors in A case, but the
beam Lorentz factor strongly decreases for B. We could not
identify a terminal Mach disk in the flow.  We find flow speeds near
(and in) the hot spot much larger than those inferred from the one
dimensional estimate. This fact was already noticed for 2D models by
\cite{KF96} and suggested by them as a plausible explanation for an 
excess in hot spot beaming.

  We find a layer of high specific internal energy (typically more
than a tenfold larger than that of the gas in the beam core, see
Fig.\,\ref{f:Aa}d) surrounding the beam like in previous axisymmetric
models \cite{AIMM99a}. The region filled by the shear layer is defined
by $0.2<f<0.95$. It is mainly composed of forward moving beam material
at a speed smaller than the beam speed (Fig.\,\ref{f:Aa}e). The
intermediate speed of the layer material is due to shear in the
beam/cocoon interface, which is also responsible for its high specific
internal energy.  The shear layer broadens with distance from 0.2$R_b$
near the nozzle to 1.1$R_b$ (in A) or 2.0$R_b$ (in B) near the head of
the jet. The diffusion of vorticity caused by numerical viscosity is
responsible for the formation of the boundary layer. Although being
caused by numerical effects (not by the physical mechanism of
turbulent shear) the properties of PPM--based difference schemes are
such that they can mimic turbulent flow to a certain degree
\cite{PW94}.  The existence of such a boundary layer has been invoked
by several authors \cite{Komiss90, Laing96} to interpret a number of
observational trends in FRI radio sources.  Such a layer will produce
a {\it top-bottom} asymmetry due to the light aberration
\cite{Aetal99b}, and additionally, it can be used to explain the rails
in polarization found by \cite{ARW99}. Other authors \cite{SBB98} have
found evidence for these boundary layers in FRIIs (3C353) radio
sources.

  
The jet's propagation proceeds in two distinct phases. First it propagates
according to a linear of 1D phase, and then the behavior depends on the 
strength of the perturbation: it accelerates to a propagation speed which 
is $\approx 20$\% larger than the corresponding 1D estimate in model A or
it deccelerates up to $0.37c$. The second result partially agrees with the 
one obtained by \cite{Netal97, Netal98}.

  The axial component of the momentum of the beam particles (integrated 
across the beam) along the axis decreases by more than a 30\% within the 
first 60\,$R_b$.
Neglecting pressure and viscous effects, and assuming stationarity the axial
momentum should be conserved, and hence the beam flow is
decelerating. The momentum loss goes along with the growth of the
boundary layer. 

In model A, although the beam material decelerates, its terminal
Lorentz factor is still large enough to produce a fast jet
propagation.  On the other hand, in 3D, the beam is prone to strong
perturbations which can affect the jet's head structure. In
particular, a simple structure like a terminal Mach shock will
probably not survive when significant 3D effects develop. It will be
substituted by more complex structures in that case, \eg by a Mach
shock which is no longer normal to the beam flow and which wobbles
around the instantaneous flow direction. Another possibility is the
generation of oblique shocks near the jet head due to off--axis
oscillations of the beam.  Both possibilities will cause a less
efficient deceleration of the beam flow at least during some
epochs. At longer time scales the growth of 3D perturbations will
cause the beam to spread its momentum over a much larger area than
that it had initially, which will efficiently reduce the jet advance
speed.

\end{document}